\documentclass{sigchi}

\toappear{This is shorter version of  deliverable D3.5  of EU project Citizen Cyberlab (Grant No 317705). Permission to make digital or hard copies of all or part of this work for personal or classroom use is   granted without fee provided that copies are not made or distributed for profit or commercial advantage and that copies bear this notice and the full citation on the first page. Abstracting with credit is permitted. To copy otherwise, or republish, to post on servers or to redistribute to lists, requires prior specific permission and/or a fee. Request permissions from poonam.yadav07@alumni.imperial.ac.uk. }

\usepackage{balance}  
\usepackage{graphics} 
\usepackage{txfonts}
\usepackage{times}    
\usepackage[pdftex]{hyperref}
\usepackage{url}      
\usepackage{color}
\usepackage{textcomp}
\usepackage{booktabs}
\usepackage{ccicons}
\usepackage{todonotes}
\usepackage{hyperref}
\usepackage{subfigure}
\usepackage[autostyle]{csquotes}

\makeatletter
\def\url@leostyle{%
  \@ifundefined{selectfont}{\def\UrlFont{\sf}}{\def\UrlFont{\small\bf\ttfamily}}}
\makeatother
\def\pprw{8.5in}
\def\pprh{11in}

\definecolor{linkColor}{RGB}{6,125,233}
\hypersetup{%
  pdftitle={SIGCHI Conference Proceedings Format},
  pdfauthor={LaTeX},
  bookmarksnumbered,
  pdfstartview={FitH},
  colorlinks,
  citecolor=black,
  filecolor=black,
  linkcolor=black,
  urlcolor=linkColor,
  breaklinks=true,
}

\begin{document}

\title{CitizenGrid: An Online Middleware for Crowdsourcing Scientific Research}

\numberofauthors{3}
\author{%
  \alignauthor{Poonam Yadav\\
   \affaddr{Imperial College London}\\
    \affaddr{London, UK}\\
    \email{p.yadav@acm.org}}\\
   \alignauthor{Jeremy Cohen\\
    \affaddr{Imperial College London}\\
    \affaddr{London, UK}\\
    \email{j.cohen@imperial.ac.uk}}\\
  \alignauthor{John Darlington\\
    \affaddr{Imperial College London}\\
    \affaddr{London, UK}\\
    \email{john.darlington@imperial.ac.uk}}\\
}
\maketitle

\begin{abstract}
In the last few years, contributions of the general public in scientific projects has increased due to the advancement of communication and computing technologies. Internet played an important role in connecting scientists and volunteers who are interested in participating in their scientific projects. However, despite potential benefits, only a limited number of crowdsourcing based large-scale science (citizen science) projects have been deployed due to the complexity involved in setting them up and running them. In this paper, we present CitizenGrid - an online middleware platform which addresses security and deployment complexity issues by making use of cloud computing and virtualisation technologies. CitizenGrid incentivises scientists to make their small-to-medium scale applications available as citizen science projects by: 1) providing a directory of projects through a web-based portal that makes applications easy to discover; 2) providing flexibility to participate in, monitor, and control multiple citizen science projects from a common interface; 3) supporting diverse categories of citizen science projects. The paper describes the design, development and evaluation of CitizenGrid and its use cases.  
\end{abstract}

\keywords{Citizen Cyber Science, Games, Volunteers, Cloud Computing, Scientists, Middleware, Distributed Computing, Crowd Sourcing}

\category{K.4.3}{Organizational Impacts}{Computer-supported collaborative work} \category{}{}{}

\section{Introduction}
In the last few years, the online scientific crowdsourcing (also known as Citizen Cyberscience~\cite{CCL})  has gained attention from educators, researchers, scientists, and policy makers due to the ongoing advances in Internet and computing technologies~\cite{Wiggins2016,CSAlliance, ECSA,CSAcademy,CSAssociation,ACSAssociation}.  The \emph{Internet-based} Citizen Science projects allow general public (volunteers) to contribute to scientific tasks or investigations from wide geographical locations. Different levels of volunteer participation have been highlighted by citizen science community~\cite{Crowdsourcing2013, Yadav2016a, Reeves2017, Kucherbaev2016, Chatzopoulos2016, Celis2016}. The two main categories are volunteer thinking and computing. In volunteer thinking, volunteers provide human computation to a project by contributing their time, effort and cognitive power to the process of solving a task~\cite{Zooniverse, Eyewire, Foldit, Eyewire2015}, e.g., the identification of ``bubbles'' in images from the Milky Way Project~\cite{MilkyWay}. However, in volunteer computing projects volunteers participate passively by donating their computing resources to perform a  project task. e.g., Seti@Home~\cite{Seti} and Test4Theory~\cite{Test4Theory}.  The volunteer computing projects requires a complex setting of server-client deployment models~\cite{Yadav2016, Dolejsova2017} and have many security and privacy concerns~\cite{BoincSecurity, Elkabbany2016, Cano2012, Bowser2017}. Due to these constraints, a large-scale volunteer computing projects that are set-up using  BOINC~\cite{Boinc} and IBM World Community Grid~\cite{WCG2015}, are only a few.  To address these issues, we build a fully-functional CitizenGrid - an online middleware platform to investigate how the use of cloud computing and virtualisation provide a feasible  and practical solution for addressing the deployment complexity and many security issues.  

In this paper, we present an overview of the CitizenGrid platform and describe its implementation, evaluation, testing processes, potential use-case scenarios  and related work along with conclusions and future work.

\section{CitizenGrid Overview} 
The CitizenGrid is designed to support these two requirements: (1) it should provide a simplified and secured online hosting platform for both volunteer thinking and computing projects; (2) it should provide user-friendly graphical interface for both scientists and volunteers and developer-friendly API (Application Programming Interface) for easy integration with other software. By taking into account these requirements, we build CitizenGrid that  supports different project deployment scenarios, user roles and functionalities to meet the design goals and guidelines of a citizen-science platform~\cite{Yadav2016a, Charlene2014}.

\subsection{CitizenGrid Supported Deployment Scenarios}

\subsubsection{Server-based applications}
Server-based applications are applications where the core application logic and all its computation runs only on server platforms that are managed by the application's operators. The servers need not necessarily be hosted by the application operators themselves and may be on the cloud or other remote servers, however, server-based applications do not push elements of their computation to end-users'(volunteers) computers. Instead, they are often volunteer thinking style applications and any interaction that is required from the end user is likely to be undertaken through a web browser where content is exchanged directly between a user's browser and the remote application server. These applications can be easier to develop, manage and operate than full volunteer computing applications. However, they lack the ability to take advantage of remote computing power on individual users' machines, something that is one of the key benefits of many volunteer computing applications. The popular citizen science platforms such as Zooniverse~\cite{Zooniverse}, Epicollect~\cite{Epicollect,Aanensen2014}  and CrowdCrafting~\cite{Crowdcrafting} support this deployment scenario and  allow data collection, aggregation and interpretation tasks.\\\\

\subsubsection{Client-server applications}
Client-server volunteer computing applications have a more traditional distributed client-server model where a client application is deployed to, and runs on, a user's local computing hardware and communicates with an application server that is run by the scientists operating the volunteer computing application. The server can send blocks of data to the client for processing and the client then sends results back to the server. Depending on the volumes of data involved and the number of potential volunteers, building such an application to be reliable and scalable, and operating it, can be a challenging task. However, these applications can take advantage of large amounts of computing power available on client PCs and servers when they would otherwise be idle, offering scientists access to huge amounts of computing power that they may not otherwise be able to access due to cost or technical constraints.  The client-server based volunteer computing projects are deployed using IBM World Community Grid~\cite{WCG2015} and use BOINC~\cite{Boinc}, middleware for distributed task management. CitizenGrid is built to support both deployment scenarios, which gives both scientists and volunteers to access and use CitizenGrid for all online citizen science projects.
\subsection{CitizenGrid Users}
\begin{figure}[h]
 \centering
 \includegraphics[width=80mm,height=60mm]{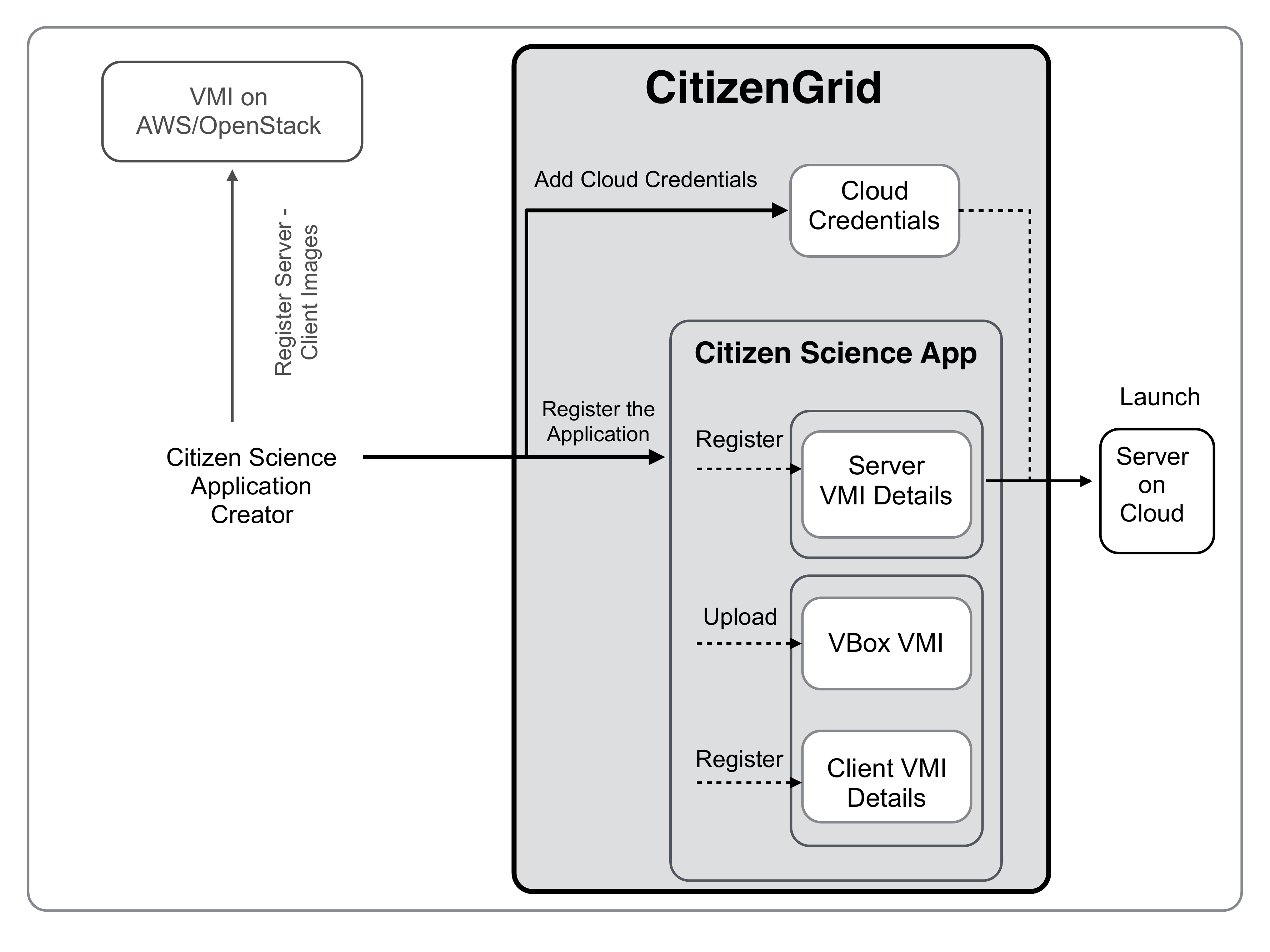} 
\caption{Scientists Interaction with CitizenGrid.}
\label{fig:CG_Scientists} 
\end{figure}

\begin{figure}[h]
 \centering
 \includegraphics[width=80mm,height=60mm]{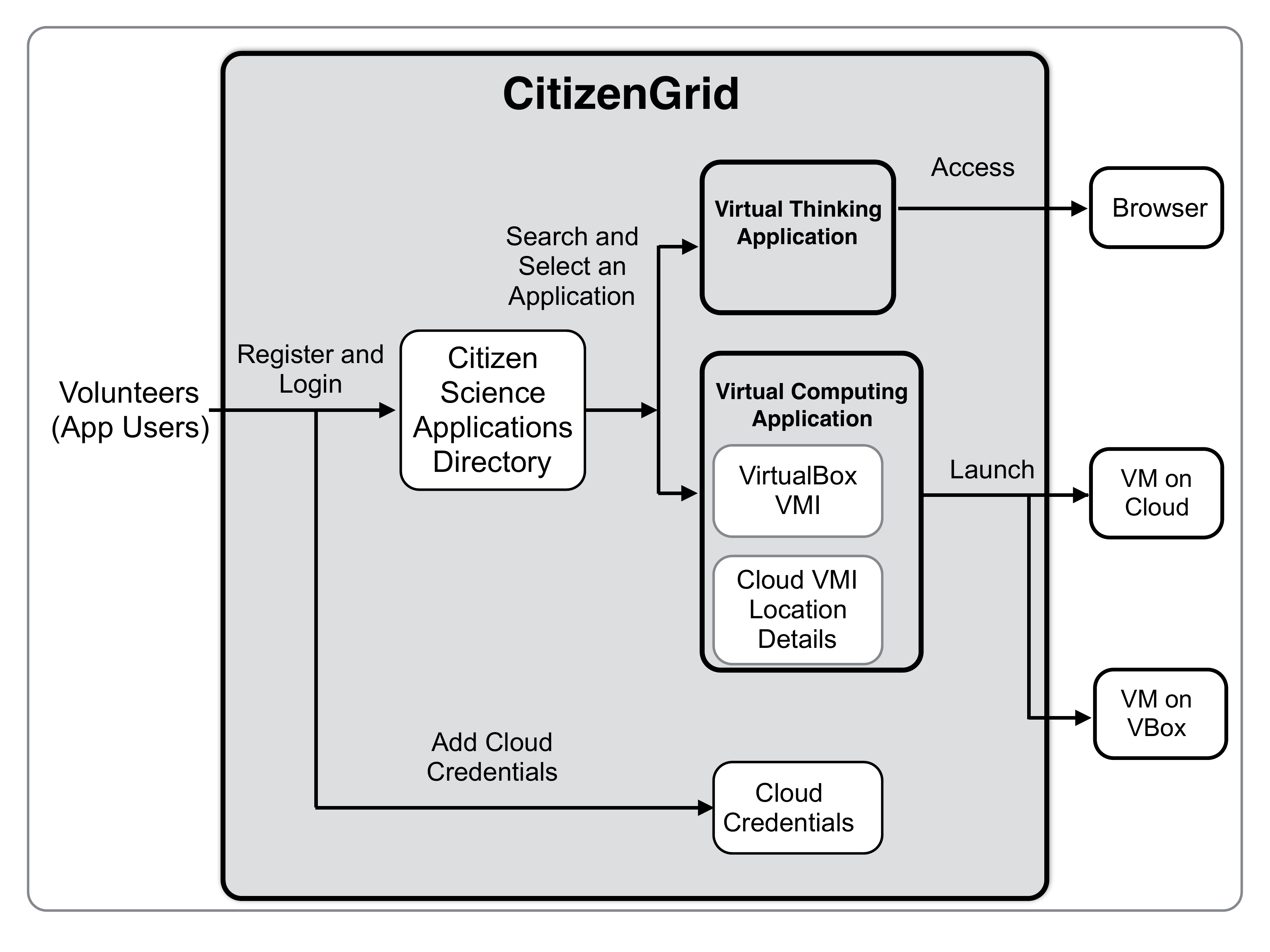} 
\caption{Volunteers Interaction with CitizenGrid.}
\label{fig:CG_Volunteers} 
\end{figure}

CitizenGrid categorises its users into two groups: (1) Application Providers and (2) Application Users:
\begin{enumerate}
\item {\textbf{Application Providers:}}  Application providers are generally entities who are responsible for setting up a citizen science project. An application provider may be either an individual scientist, a team or an organisation. The application provider hosts their Citizen Science application on the CitizenGrid platform so that it is publicly visible and available to members of the general public to participate in. Figure~\ref{fig:CG_Scientists} illustrates application providers'/scientists'/creators' interaction with the CitizenGrid application.
\item {\textbf{Application User:}} Application users are volunteers / citizens / members of the general public who are interested in online citizen science applications. They can look at the available applications via the CitizenGrid website and volunteer to take part in a project, selecting their chosen application and following the instructions to set up/work with it. Figure~\ref{fig:CG_Volunteers} illustrates the process of interacting with CitizenGrid for application users.
\end{enumerate}

\subsection{CitizenGrid Functionality}
The CitizenGrid web application provides a variety of functionality for both end-users who want to take part in online citizen science projects and the creators and operators of those projects who want to host their project(s) and attract volunteers to take part. In this section we provide an overview of the functionality of the CitizenGrid web application along with a number of screenshots to demonstrate what an individual interacting with the application sees in their browser. 

\subsubsection{User Management}

The CitizenGrid web application provides a standard sign-up approach for new users.  An individual who has registered with the CitizenGrid application can be both an application provider and an application user. However, both user types see slightly different user interfaces that are tailored to specific role requirements so it is necessary to switch between roles depending on the task that a signed in user wishes to undertake. The application enables registered and signed-in users to manage their account by updating user details as required and there is a facility to recover/reset a forgotten password.  
CitizenGrid provides group management facility that allows end users to create a group, join an existing group, leave a group and delete a group if the user is the owner of the group.   

\subsubsection{Application Management}
\begin{figure}[h]
 \centering
 \includegraphics[width=80mm,height=60mm]{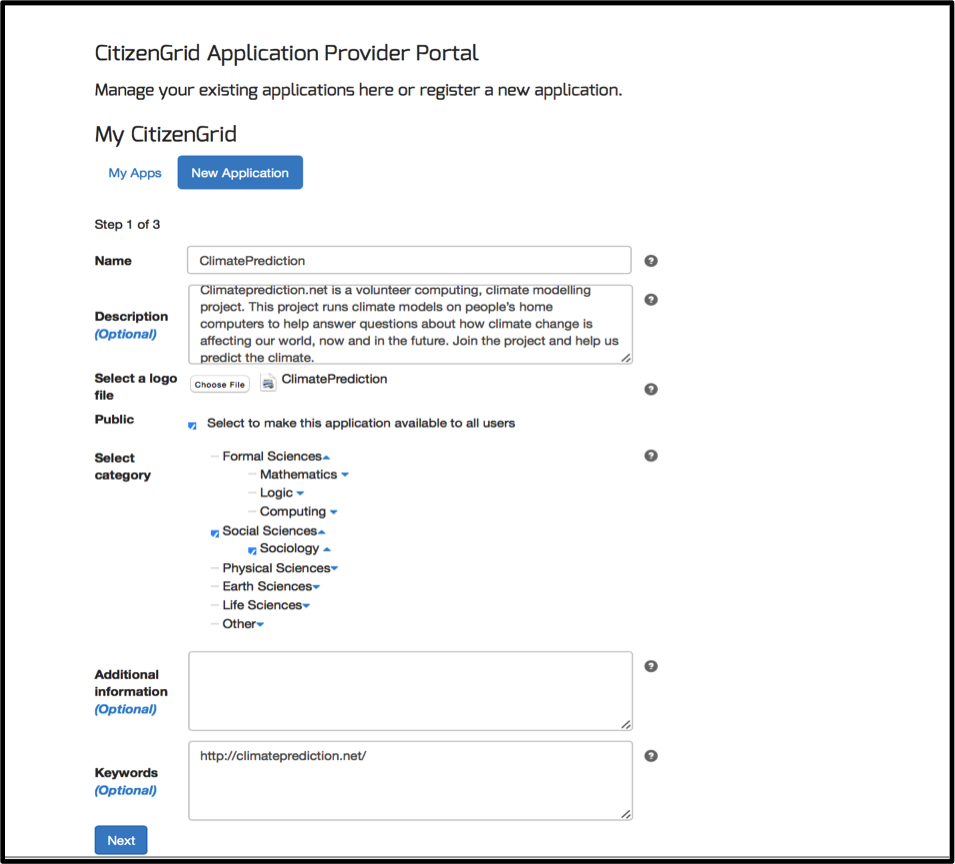} 
\caption{CitizenGrid New Application Creation Interface.}
\label{fig:CG_App} 
\end{figure}

A user who is an application provider can add their applications to CitizenGrid via the ``New Application'' form (see Figure~\ref{fig:CG_App}).

This form allows an application provider to specify a name and description for their application, assign it to a particular category and add keywords to help application users to find the application via CitizenGrid's application search feature. Applications are tagged to specific subject area categories so that they can be recommended to users who are interested in volunteering to take part in projects in specific scientific fields. 

\begin{figure}[h]
 \centering
 \includegraphics[width=80mm,height=60mm]{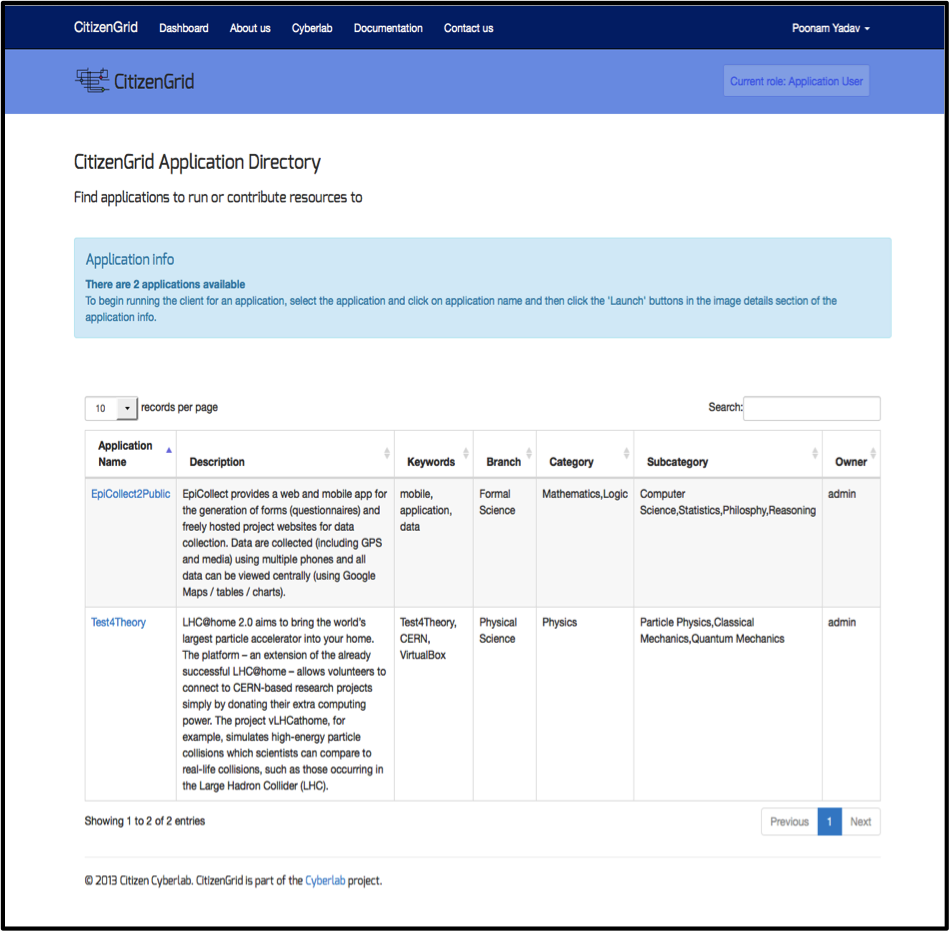} 
\caption{CitizenGrid Application Directory.}
\label{fig:CG_public_app} 
\end{figure}

\begin{figure}[h]
 \centering
 \includegraphics[width=80mm,height=60mm]{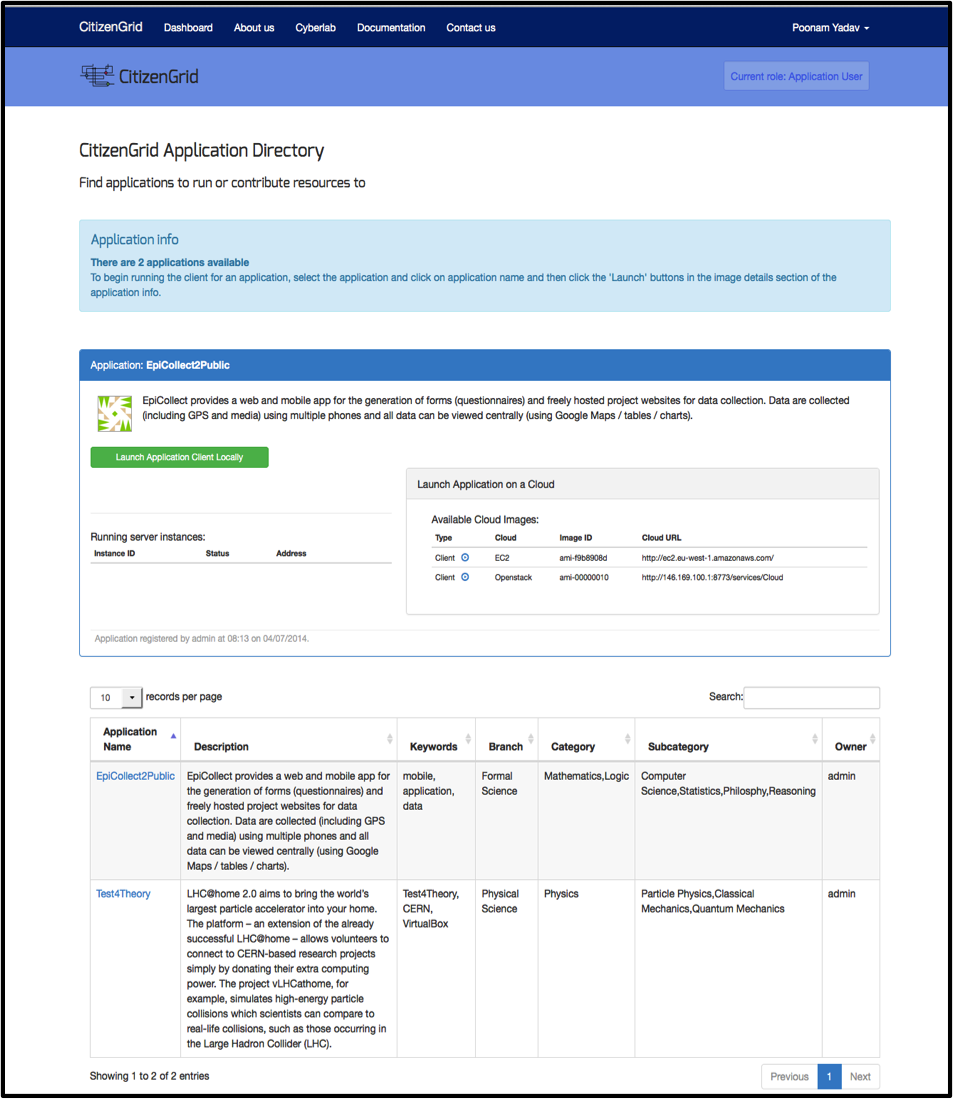} 
\caption{CitizenGrid Application's Detail View.}
\label{fig:CG_public_apps} 
\end{figure}

Where a volunteer computing application is being registered, it is possible to upload a machine image for an application client that can be deployed to volunteers computers. At present, upload of VirtualBox images is supported but an application provider can also provide details of a cloud computing image for an OpenStack cloud or the Amazon Web Services EC2 public cloud which could enable volunteers to fund the operation of a cloud computing server to undertake computation for the project. Application server images can also be registered to support the case where the application provider wants to run the server-side of their application on a remote cloud platform. Once an application provider has registered an application it will appear in their \enquote{My Apps}  list where they can manage their existing applications, for example, by adding or removing images.
CitizenGrid allows users to see either a list view (Figure~\ref{fig:CG_public_app}) or expanded view (Figure~\ref{fig:CG_public_apps}) of any selected application and search the available public applications using free text search. Users can list available applications in alphabetical ascending or descending order by clicking on the vertical arrows shown in the column headers. The columns are: application name, description, keywords, branch, category, subcategory, and owner.

\begin{figure}[h]
 \centering
 \includegraphics[width=80mm,height=60mm]{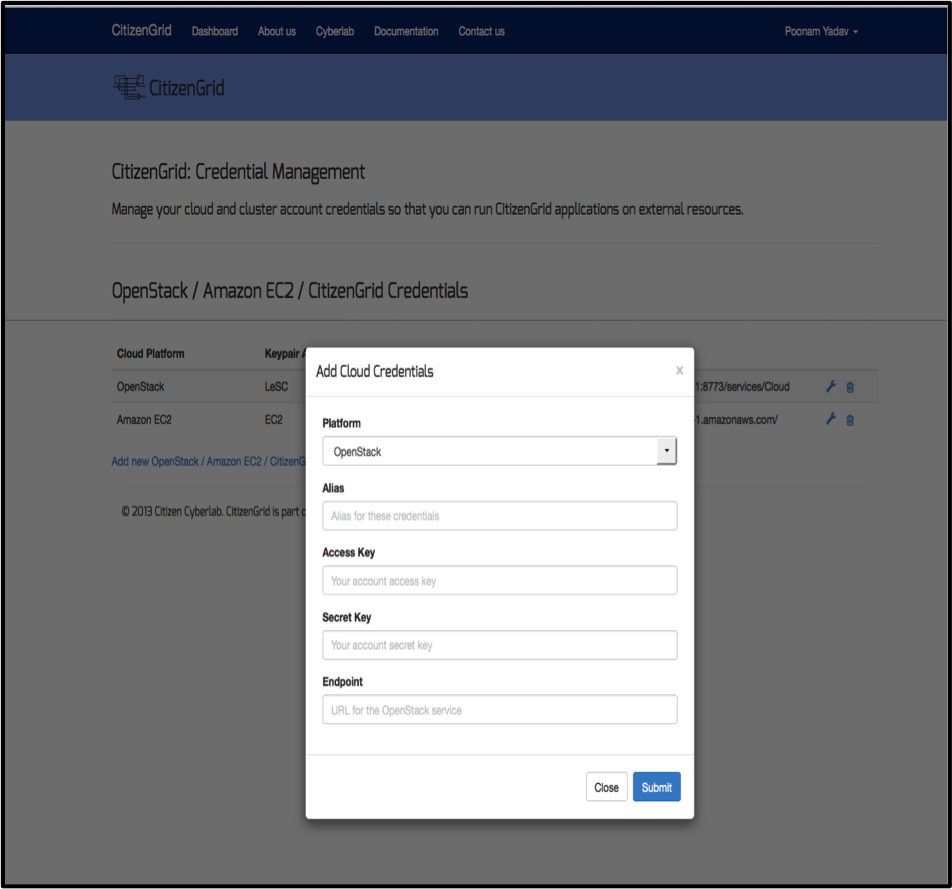} 
\caption{CitizenGrid Cloud Credential Form.}
\label{fig:CG_cloud} 
\end{figure}
\subsubsection{Application Deployment on Cloud Platforms}
\begin{figure}[h]
 \centering
 \includegraphics[width=80mm,height=60mm]{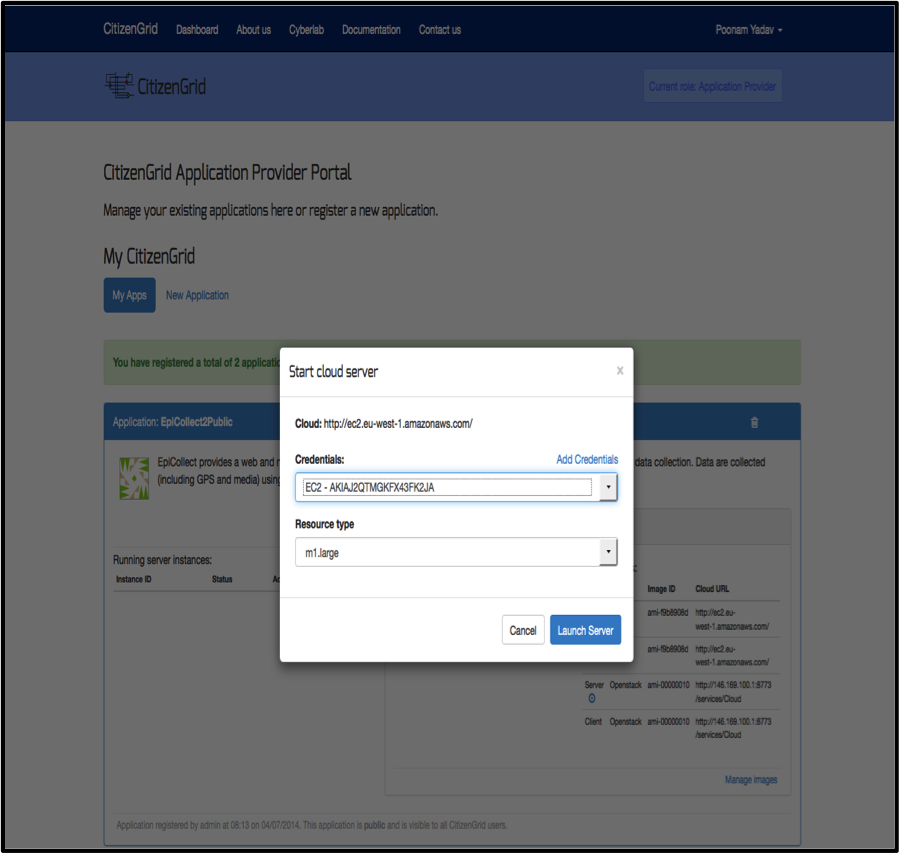} 
\caption{CitizenGrid Application Server Launch.}
\label{fig:CG_applaunch} 
\end{figure}

CitizenGrid provides support for running server-side and client application images on the cloud. 
In its current implementation, CitizenGrid supports Amazon EC2 and OpenStack clouds. This functionality requires developers to first upload and register their application images on each cloud platform that they wish to use to run application servers. Once an image is registered, the application provider registers their cloud platform account details with CitizenGrid. The application provider sets up their server image using the Manage Images menu option by providing CitizenGrid with the unique identifier assigned to their image by the cloud platform. The application server image now appears alongside the application details. To run an instance of the application's server on a cloud resource, the application provider clicks the play symbol that is displayed with the image details. It opens a launcher form as shown in Figure~\ref{fig:CG_applaunch}, which shows a list of cloud credentials that are available for the specific cloud platform that the image is registered with. Once the user clicks on the launch server button, and the server start-up process has begun, the \enquote{Running server instances} section of the application information shows the status of the server. A running cloud server can be terminated by clicking the shutdown \enquote{X} button alongside the instance status. 

\begin{figure}[h]
 \centering
 \includegraphics[width=80mm,height=60mm]{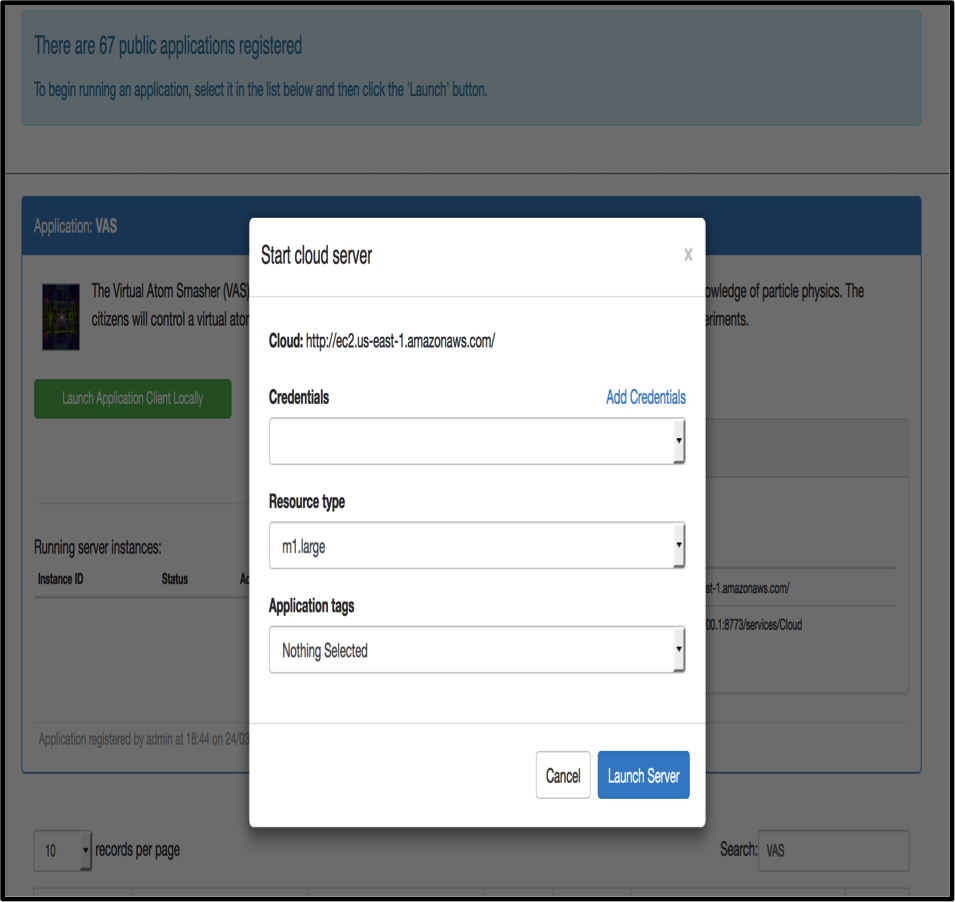} 
\caption{Application Client Launch on the Cloud.}
\label{fig:CG_launch_client} 
\end{figure}

\begin{figure}[h]
 \centering
 \includegraphics[width=80mm,height=60mm]{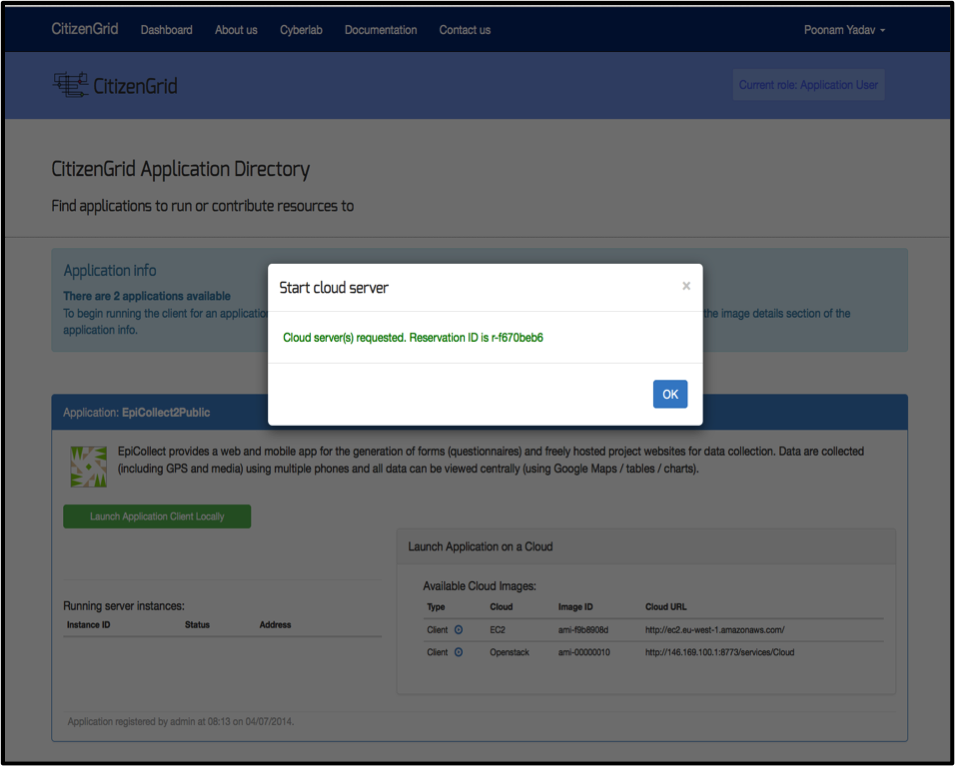} 
\caption{Application Client Launched successfully on the Cloud.}
\label{fig:CG_launch_client1} 
\end{figure}

\begin{figure}[h]
 \centering
 \includegraphics[width=80mm,height=60mm]{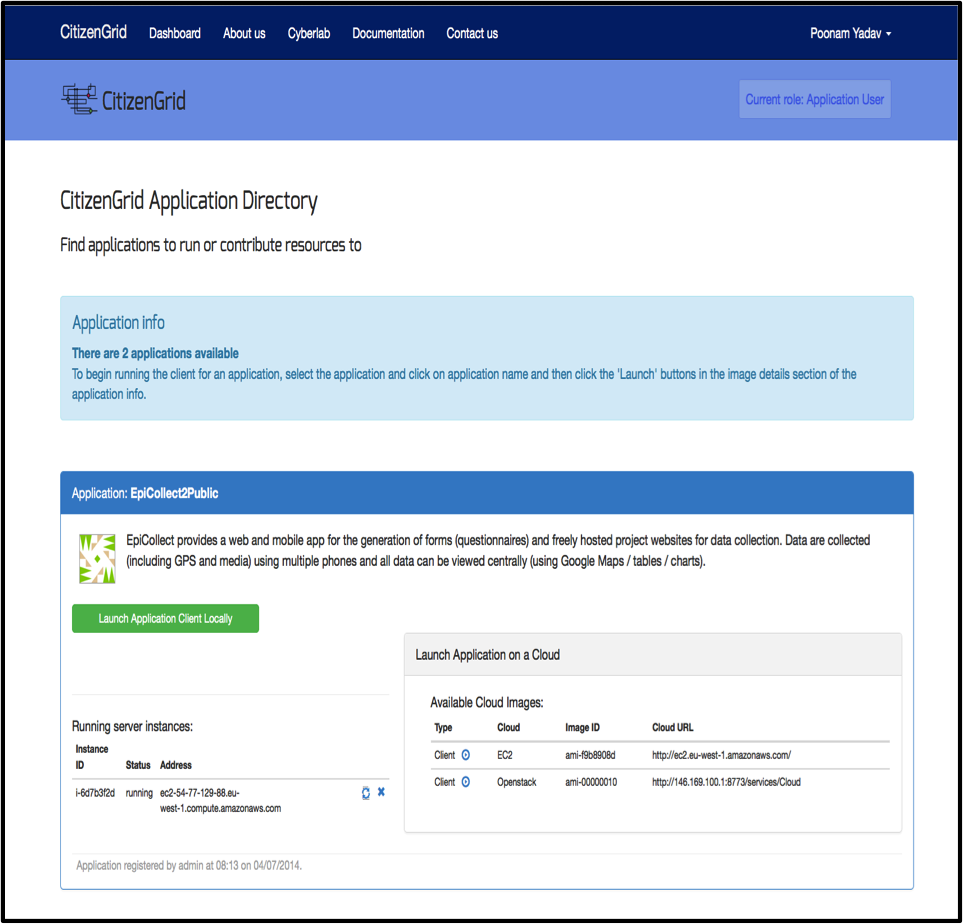} 
\caption{Application Client Running Status shown on the dashboard.}
\label{fig:CG_Clientrun} 
\end{figure}

This feature gives CitizenGrid users (volunteers) flexibility to choose between running an application client on their local resource or public clouds such as Amazon EC2 or an OpenStack private cloud. Before launching an application client  on the cloud, the user has to register their cloud credentials with CitizenGrid as shown in Figure~\ref{fig:CG_cloud}. The user clicks on the dashboard to see the available public applications.  When the user clicks on the application, the expanded application view opens in the same page.  In the application description, there is a block on the right-hand side column with the title \enquote{launch application on the cloud}. If there are any entries in this table, it means the application's client images are registered on the cloud platforms mentioned in the column named \enquote{cloud}. The \enquote{type} column shows whether the entry represents client or server parts of the application. This entry will always be set to \enquote{client} when viewed by application users.  By clicking on the \enquote{play icon} next to the client image entry, the user can start the process of launching the application client on the cloud. A new window opens up as shown in Figure~\ref{fig:CG_launch_client}, which asks users to select the cloud credentials for the specific cloud account they would like to use from the drop-down selection list. The volunteers also select the instance resource type -  an identifier that determines the specification of the resource started on the cloud platform. For EC2, the values are m1.large and m1.small.  Volunteers can obtain details from the application's properties about the type/specification of resources they should choose when running a cloud-based client. Clicking launch sends a launch request with the additional details to the cloud service. If the request is successful, a reservation id is returned and displayed in a new pop-up window (Figure~\ref{fig:CG_launch_client1}), which, can be closed by clicking on the OK button. Details of the running instance(s) will then appear in the application details panel providing the instance ID, instance status, and the IP address assigned to the cloud instance (Figure~\ref{fig:CG_Clientrun}).  An instance can be stopped by clicking on the \enquote{X} icon to the right of the instance details.

\subsubsection{Application Deployment on Volunteer's Computers}
\begin{figure}[h]
 \centering
 \includegraphics[width=80mm,height=60mm]{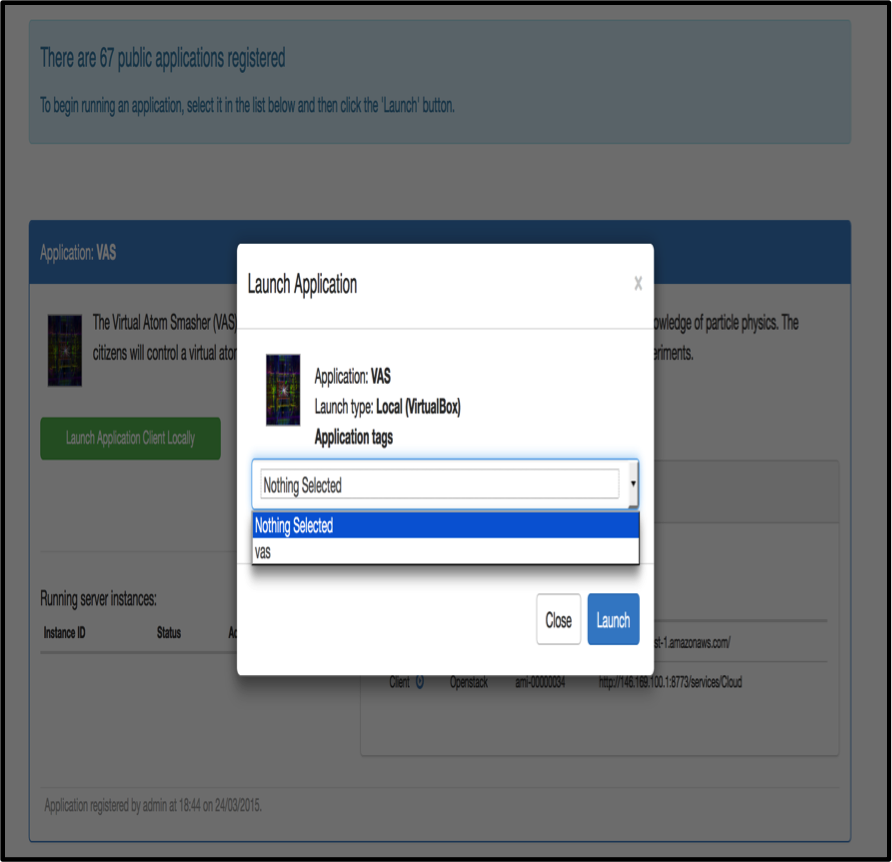} 
\caption{CitizenGrid Application Local machine Launch View.}
\label{fig:CG_launch_view} 
\end{figure}

\begin{figure}[h]
 \centering
 \includegraphics[width=80mm,height=60mm]{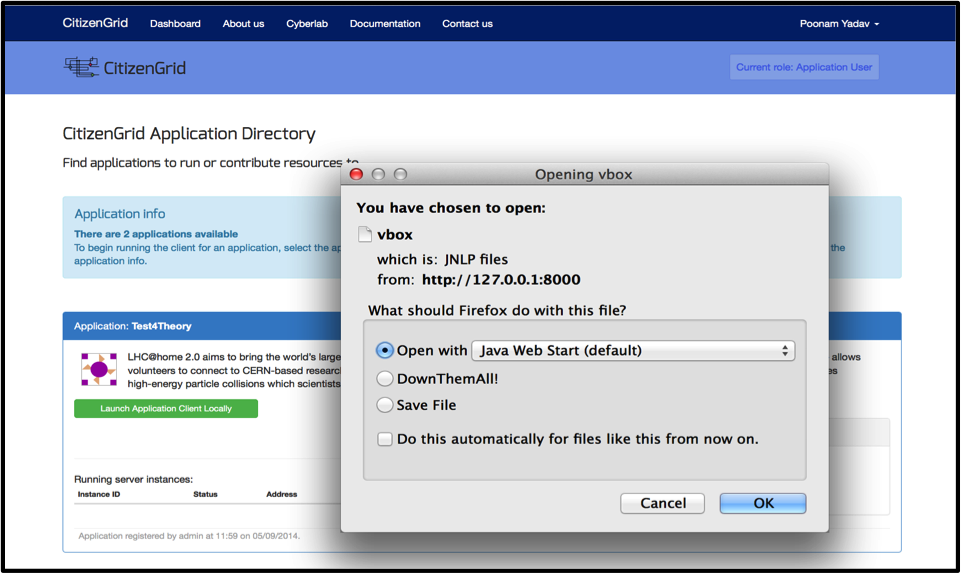} 
\caption{CitizenGrid Application Local machine Launch View.}
\label{fig:CG_launch_view1} 
\end{figure}

CitizenGrid provides the ability for end users (in the application user role) to run volunteer computing application clients on their local system through VirtualBox. To launch an application, the user clicks on the relevant \enquote{Launch Application Client} button.  It opens up the \enquote{Launch application} form (Figure~\ref{fig:CG_launch_view}); when the user clicks on the \enquote{launch} button (see Figure~\ref{fig:CG_launch_view1}), the client is launched on the user's local system in a virtual machine run via VirtualBox.  The client downloads the applications' client VirtualBox image only once.  The running instance then directly connects to the project server  to download computing tasks and returns processed results directly to the server. Hence, providing a secure environment by encapsulating  citizen science application  and isolating it from the volunteers'  personal data and processes.


\subsubsection{Public RESTful API}
The CitizenGrid platform provides a public RESTful API (application programming interface).  The API allows CitizenGrid to be integrated with other platforms via its REST interface. The CitizenGrid API allows programmatic access to the full functionality of the CitizenGrid system through a series of operations, their inputs and outputs, and underlying types. 

\section{CitizenGrid Implementation}\label{sec:implementation}
The CitizenGrid platform has been designed and developed based on the Django Framework~\cite{Django}. Django is a free and open-source high-level Python web framework, which follows the model-view-controller (MVC) architecture pattern. The  Django is Web Services Gateway Interface (WSGI) compatible and is therefore able to operate behind any WSGI-compliant servers such as Apache and NginX~\cite{Nginx}. 
\begin{figure}[h]
 \centering
 \includegraphics[width=80mm,height=60mm]{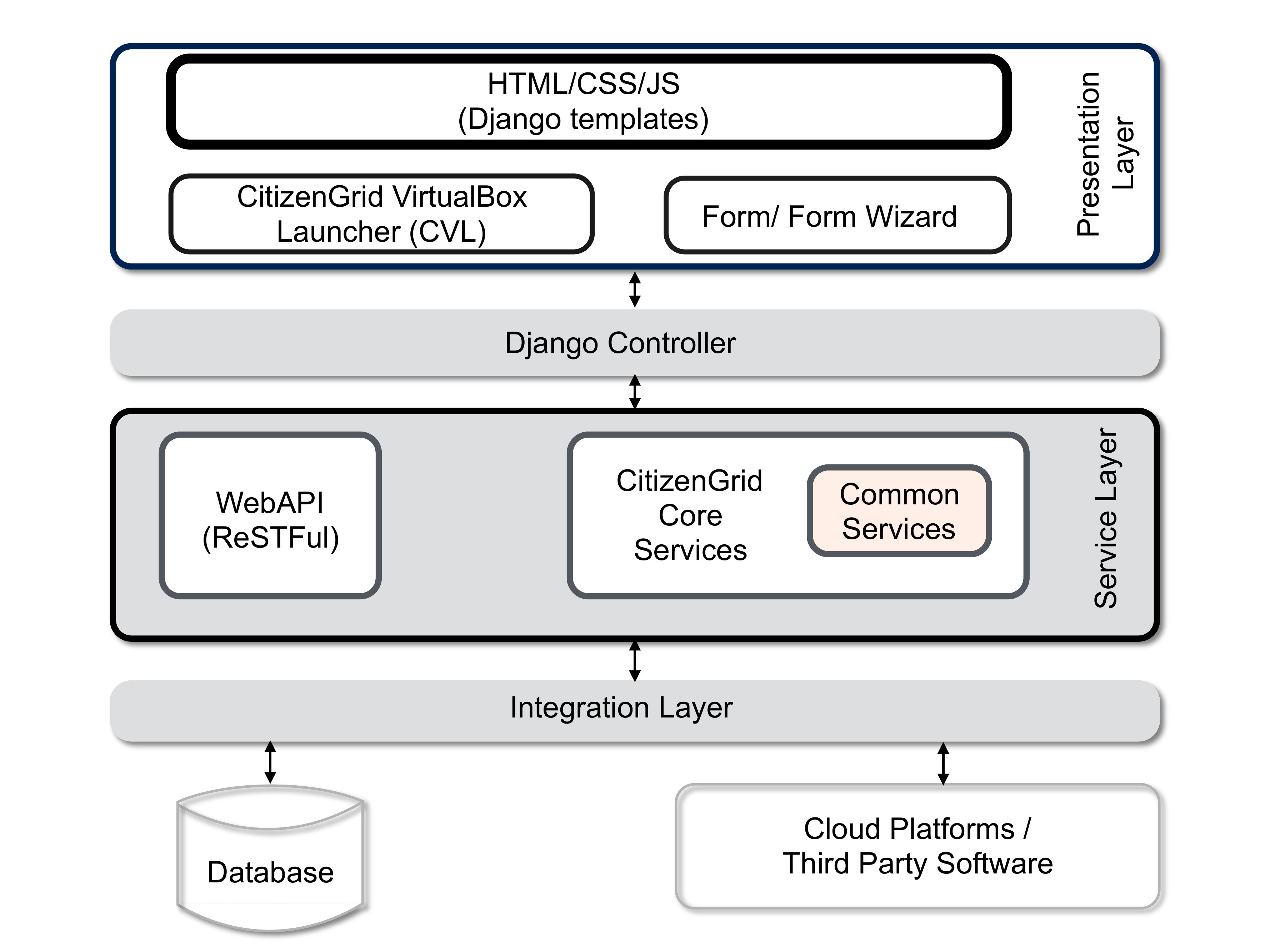} 
\caption{CitizenGrid Architecture.}
\label{fig:CG_Architecture} 
\end{figure}
In Figure~\ref{fig:CG_Architecture}, we present the CitizenGrid's frameworks layered architecture. This layered architecture is built on top of the standard MVC model of Django which defines the general structure of the code consisting of a data model, views which describe presentation of information to the user and a controller layer which handles the logic of managing data and making this available to the views. The CitizenGrid application design can be categorised into 5 layers: Presentation Layer, Application Service Layer, Common Service Layer, Integration Layer, Web API. We now provide an overview of each of these five layers. The interaction of layers among layers is shown in Figure~\ref{fig:CG_Architecture}. \\\\
\subsection{Presentation Layer}
The presentation layer handles presentation of information to the user and user interaction with the system. It is tightly coupled to the view layer of the MVC structure. Making use of Django's presentation philosophy, CitizenGrid views are driven using templates. A Django template is an HTML file containing additional Django-specific markup that defines content that is dynamically generated and included into the template in response to a user request.  Django has a rich template language that allows advanced management of data within templates. CitizenGrid forms for taking inputs from users are designed using Django's built-in form functionality and FormWizard. The presentation layer uses the following web technologies and libraries to make the views available to users via a web browser - (a) JQuery, a JavaScript library supporting a wide-range of web programming requirements; (b) Bootstrap JS and CSS framework is used as a visual framework for web-page design; (c) Django Form wizard Framework  for designing form wizards; (d) Font-Awesome CSS library for providing font effects.\\\\

\subsection{Application Service Layer}
CitizenGrid's Application Service layer implements core logic for the CitizenGrid use cases.  Combined with the Common Services Layer, this provides the main implementation logic for the CitizenGrid framework, acting as an intermediary between the data model and the presentation layer. The services are realised using Django's view framework, implemented in Python code that obtains data from the data store and processes and prepares this data before passing it to the templates used in the presentation layer.  This layer communicates with Django's Object-Relational Mapping functionality through Model Views that handle database queries. Communication with external cloud platforms is done through the integration layer. 
In our implementation, we organise views into five general groups, as follows:\\
\textbf{Standard Views:} These views handle requests, which are not secured (i.e. no user login is required to access these views). For example, a request for the CitizenGrid homepage, \textit{contact us} page, or to view the documentation is handled by a standard view.\\
\textbf{Secure Views:} These views require a user to be logged in in order to access them. Requests from authenticated users interact with the presentation layer and the CitizenGrid database through the integration layer. These views are implemented using both class-based views and function-based views, two different styles of view defined by the Django framework. 
For authenticated users, the CitizenGrid platform maintains two separate views based on a users currently selected role  either application provider or application user. For example, when an authenticated user who has their role set as \textit{application provider} clicks on the \textit{dashboard} tab, they see the application provider view specific to them, containing details of their registered applications and associated configuration. \\
\textbf{Launch views:}  These views handle requests from authenticated users and interact with the presentation layer, CitizenGrid database and external cloud platforms through the integration layer. They manage the process of running citizen science application clients or servers on third-party cloud resources.\\
\textbf{API views:} CitizenGrid provides a RESTful API. The API views provide the same functionality as the secure views and launch views but rather than returning templates to be processed and sent to a web browser, data is returned in a format such as JSON or text, in order to be programmatically processed by a computational client. These views handle requests from authenticated users and interact with the presentation layer and CitizenGrid database through the integration layer. As these views provide the same functionality as the secure views and launch views, duplicating code for these views is not considered good program-ming practice, especially in the Django framework model.  Attempts have therefore been made to keep duplication of code to a minimum and ensure as much sharing of functionality as possible. \\
\textbf{Utility views:} In order to adhere to Django's DRY (don't repeat yourself) principles, we have refactored some of the views in order to aggregate common functionality into a set of utility views.  These views are not directly mapped to URLs but are used directly by secure, launch and API views.
\subsection{Common Services Layer}
As the name suggests, the Common Services layer implements common concerns and justifies the DRY principle. It also realises some of the utility concerns of the CitizenGrid application. The common services are used by more than one view or function. These include security and encryption/decryption services   and other functions generally defined as utility functions. In the CitizenGrid implementation, we use the \textit{Crypto.Cipher} package for implementing the encryption and decryption mechanisms for secure information such as pass-words and private keys before storing them to the database and after retrieving from data
\subsection{Integration Layer}
This layer decouples the above layer from interacting with any third party services e.g. interaction with AWS and OpenStack components.  In CitizenGrid, the integration layer controls the following interactions:
Interaction between CitizenGrid and the AWS EC2 /OpenStack cloud platforms:  
\begin{itemize}
\item [--] AWS EC2 and OpenStack provide a secured API for accessing their services. In order to use these services in an application such as CitizenGrid, developers have two options. One is the direct use of the cloud platform API directly from the application's code, the other is the use of a higher-level library that abstracts away some of the complexities of using the cloud APIs directly. In particular, such services can be very useful in hiding differences, which may be subtle, between the APIs of different cloud platforms. In the case of CitizenGrid, access to the cloud platforms is carried out in the launch views. We use the generic Boto~\cite{Boto} Python library, which provides sup-port for accessing AWS EC2, OpenStack, and other cloud platforms.
  
\item [--] Interaction of CitizenGrid with VirtualBox: 
Launching VirtualBox images that are hosted on the CitizenGrid platform to a user's local machine, requires three steps: 1) Download the VirtualBox image onto the local machine. 2) Connect the image in VirtualBox via the VirtualBox Web Service API and 3) Start the virtual machine image in VirtualBox. The CitizenGrid platform uses Java Web Start, receiving a JNLP (Java network launch protocol~\cite{JNLP}) file that triggers the download of a Java Web Start application~\cite{JWST}, which starts 
 CVL (CitizenGrid VirtualBox Launcher). CVL downloads the VirtualBox VM image(s) associated with the application that the user wishes to run and configures VirtualBox to launch the VM. 

\item [--] Interaction with Database:  For interacting with the CitizenGrid database, we use Django's built-in DB backend support.  Django's built-in database API provides an abstraction layer to perform database operations, including the Object-Relational Mapping (ORM) process of converting between data in the relational structure stored in the database and the Python objects that the developer interacts with in the Django application. As part of this functionality, Django handles creation of database tables from \textit{model classes} written by the application developer.  An instance of a custom model class represents a record in the table. The database abstraction layer gives developers a great deal of flexibility and significantly simplifies the process of interact-ing with the database.  We used the django-south package for assisting with data-base schema migration. In CitizenGrid development  and testing environment we used SQLite3 database, whereas in our production level deployment, we configured CitizenGrid to use MySQL database.
\end{itemize}

\subsection{Web API}
This layer contains the REST API, which can be used by registered users to interact with the CitizenGrid platform programmatically. These API are built using the Django REST framework [20] and are secured using the OAuth2 library.  For serializing and de-serializing the complex model datatypes into Python data types, we used class based Serializers, which extend Django REST framework's ModelSerializer and HyperlinkedModelSerializer classes. The Python datatypes are easily rendered to requested content types (JSON, XML, Text, etc.) using the REST framework's classes, including JSONRenderer and BrowsableAPIRenderer.

\section{Testing and Evaluation}\label{sec:evaluation}
The CitizenGrid platform has been tested and evaluated following each development cycle. The categories we looked at while testing were functionality, usability, compatibility, security, database integrity and performance~\cite{WebTest, TestChecklist, Matera2006} and performed tests in each category during the development  and testing phases. 
\subsection{User Tracking and Platform Analytics}
Volunteer participation analytics, such as frequency of login to the platform and participation in any particular application, will help scientists to understand how users learn and participate with the registered applications. Information, such as the number of users participating at any given time or from a given geographic location, will provide interesting insights into citizen science and the way that the CitizenGrid platform is being used.  To obtain the analytics, we use both server-side and client-side tracking mechanisms. On the server side we gather the following data analytics variables:
\subsubsection{Platform-wide analytics}
CitizenGrid platform collects statistical data points, for example, total number of (active) projects that are registered on the CitizenGrid at a given point of time, total number of registered users, total number of projects under different branches, categories and subcategories, popular projects defined by highest degree of participation.  This data provides interesting co-relation between users' preferences and citizen science project features.

\subsubsection{Individual application user analytics}
\begin{figure}[h]
 \centering
 \includegraphics[width=80mm,height=45mm]{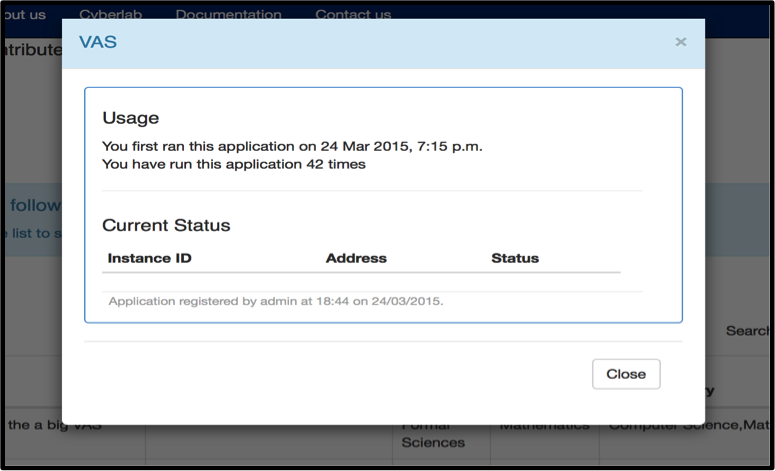} 
\caption{Individual User participation record in a particular application.}
\label{fig:partrecord} 
\end{figure}

CitizenGrid  platform provides the volunteers'  the detail analytics of the applications in which they have been participating. This includes - participation history  in the dashboard under the \enquote{My Apps} tab shows a list of applications in which the user has participated as a volunteer.  A detailed view of the application history can be viewed by clicking on the application name, which opens up a usage box that shows when the user first ran the application and how many times it has been run since then (see Figure~\ref{fig:partrecord}). Volunteers can also see the status of currently running applications, for example,  details of the running instance of the application with Instance ID, Address and Status.

\subsubsection{Platform-wide User Tracking}
\begin{figure}[h]
 \centering
 \includegraphics[width=80mm,height=40mm]{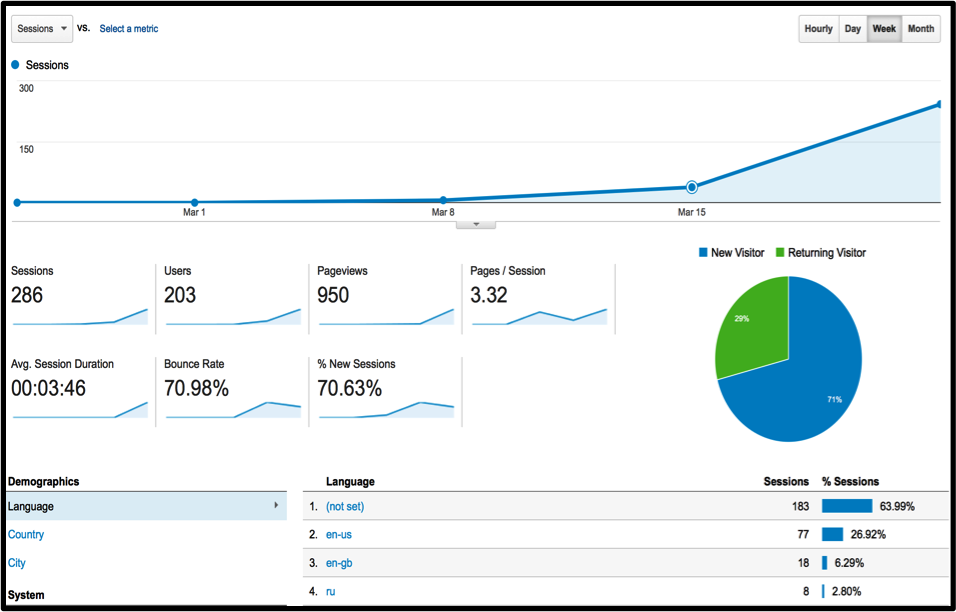} 
\caption{Page view  tracking analytics of CitizenGrid.}
\label{fig:googleanalytics} 
\end{figure}

For client-side user tracking and behaviour, we are using the Google-tag manager and Google-web-analytics. Google analytics provides a rich API for implementing detailed user tracking.  It provides event tracking, in-page analytics, and site speed analytics, which is very helpful in obtaining the details of the CitizenGrid platform usage from the users' perspective.
We use \enquote{event tracking} when a user searches for an application, clicks on the application to see the details, clicks on the launch button to launch an application locally, clicks on the run icon for launching an application in the cloud, and clicks on the stop button to stop a cloud instance.
By making use of in-page view tracking, we collect a detailed overview of sessions, users, page-views, pages/sessions, average session duration, bounce rate, and the number of new visitors/ returning visitors) of the CitizenGrid production server. The graph shown in Figure~\ref{fig:googleanalytics} presents the in-page view tracking  analytics of  the CitizenGrid. 

\subsection{Performance Testing}
\begin{figure}[h]
 \centering
 \includegraphics[width=80mm,height=40mm]{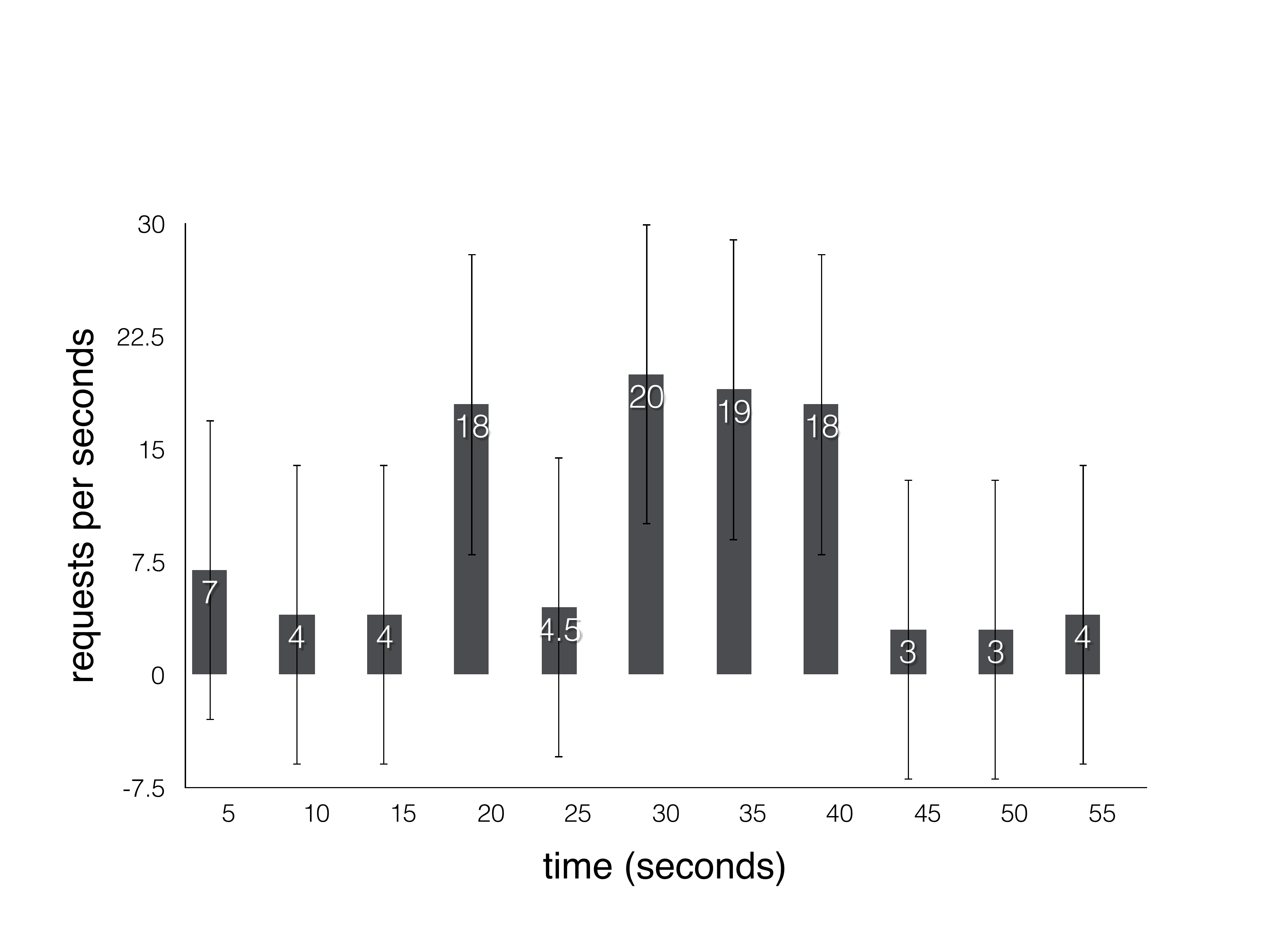} 
  \includegraphics[width=80mm,height=40mm]{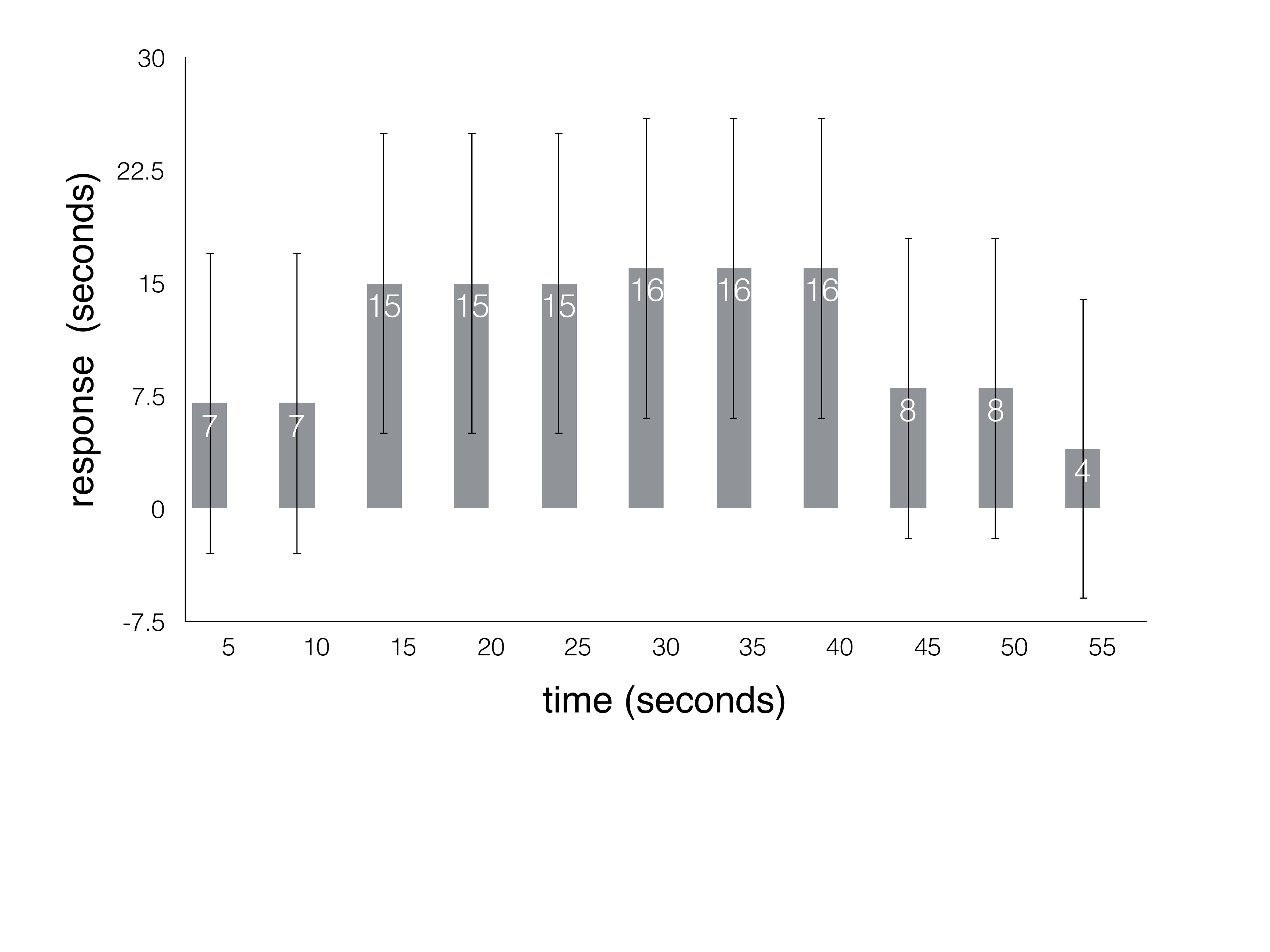} 
 \caption{Load-test statics for CitizenGrid server - 2000 concurrent users.}
\label{fig:locust1} 

\end{figure}

The website has been tested for varying levels of load (concurrent users and swarms of users) using an open source tool - Locust~\cite{Locust}. The test scenarios are written as a locust file (python) and simulated for execution by hundreds of concurrent users.  Locust is completely event-based and helped us simulate thousands of concurrent users on a single machine. The performance testing was simulated for both secure and insecure urls. The statistics for the performance tests are illustrated in Figure~\ref{fig:locust1}. The results show that in high-load scenarios where every user is making nearly 20 requests/s,  the response time is around 16 seconds which involves setting up download files for virtual machines.

\section{Use-case Scenarios}\label{sec:usecases}
\subsection{Virtual Atom Smasher (VAS)}
The Virtual Atom Smasher is an online volunteer computing game~\cite{VAS2015, VAS2015a}. It is an open-source volunteer computing based science gaming platform~\cite{VAS}, which allows users to contribute to the scientific discoveries at CERN without any prior knowledge of particle physics. A close collaboration between the VAS and CitizenGrid developers made it possible to connect the new game and CitizenGrid in an effective way~\cite{Yadav2017CGVas}. 
\begin{figure}[h]
 \centering
 \includegraphics[width=80mm,height=40mm]{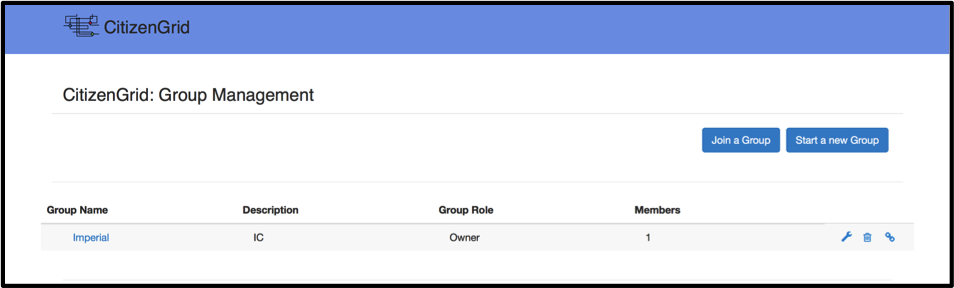} 
\caption{User Group Management Interface on CitizenGrid.}
\label{fig:cggroup} 
\end{figure}
In VAS, game players control a virtual atom collider that produces scientific data analogous to those generated by real LHC experiments. Their goal is to tune the parameters of the virtual collider until the simulated results most precisely match the experimental ones.  Each parameter change in the game console by the game player requires extensive simulation. Game players use their local machine through CitizenGrid to run these simulations, for which they first need to install VirtualBox and run downloaded VirtualBox images. If game players are competing with other players and want to run simulations faster, and their local machine is not sufficient, CitizenGrid can behave as a middleware platform, allowing users (game players as well other volunteers) to launch VAS client images on their computers and on the cloud as virtual machines. Thus, in this case the CitizenGrid platform acts a computing resource manager for VAS. 
CitizenGrid's  group management feature is very useful for the VAS players as well as for volunteers who want to donate their computing resources to their favourite game player/teams.  CitizenGrid not only manages groups but also dynamically configures a VAS worker node based on their team selection. This function is particularly useful for VAS, where each simulation submitted by a user requires a lot of processing power to run. CitizenGrid Groups allow institutions to donate computing resources to specific users to help them play the game.

\begin{figure}[h]
 \centering
 \includegraphics[width=80mm,height=40mm]{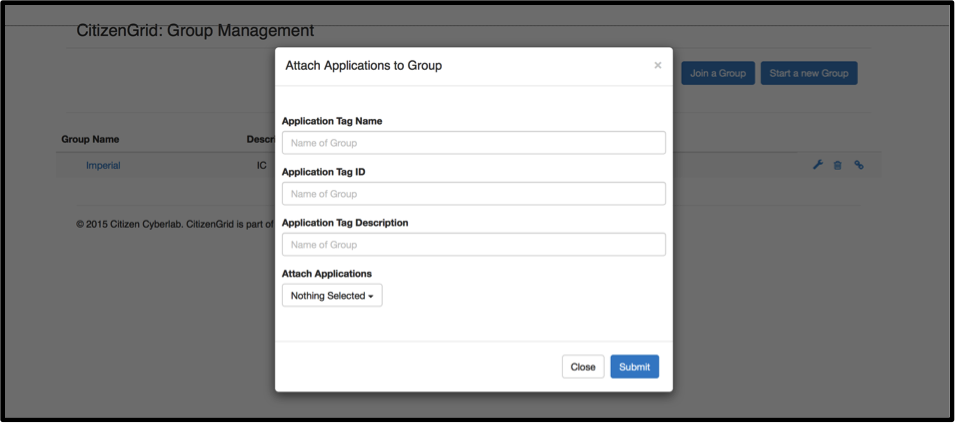} 
\caption{Creating a link between a application and a user group.}
\label{fig:vasid} 
\end{figure}
Figure~\ref{fig:cggroup} shows the group management interface, where users can attach an application to their selected group. In Figure~\ref{fig:vasid}, as an application Tag id, the group owner provides the queue id of a VAS worker. In the current version of VAS, a queue id represents a team of players.  They can attach a VAS worker node to a queue id. The next time any member of the group launches a VAS worker node, they have an option of selecting the queue id available for their group. 
\subsection{GeoKey}
GeoKey is a platform for participatory mapping. It provides communities with a web-based infrastructure to collect, share and discuss local knowledge~\cite{GeoKey}. It is an open-source platform and users can set up their own mapping project via GeoKey's administration, by creating custom data structures that support data collection and validation for specific use-cases. The CitizenGrid platform allows users to launch their own instance of GeoKey, providing support for them to launch and manage more than one instance on different clouds and to manage other servers as well.  
\section{Related Work}\label{sec:relatedwork}
There are a number of client-server based volunteer computing middleware platforms that have been developed in last two decades, e.g., BOINC~\cite{Boinc}, Co-pilot~\cite{Copilot}, Cosm~\cite{Cosm1995}, and Bayanihan~\cite{Luis1999}.  The first few successful projects, for example, Folding@home~\cite{Folding2000,Beberg2009} and Distributed.net~\cite{Distributednet} use the Cosm networking libraries for distributed networking. However, the majority of volunteer computing projects that have started in last few years use BOINC~\cite{Boinc}. BOINC is a distributed batch processing system. BOINC clients, which run on volunteers' computers, can pull tasks from the BOINC server when they are idle in order to process them locally. The volunteer's computer stores a batch of processed tasks before sending them back to the project server.  Co-pilot~\cite{Copilot} is another distributed task management system that is designed for CERN volunteer computing projects, which acts as a \textit{gateway} between CERN's grid-infrastructure and volunteer resources. Co-pilot allows the distribution of CERN worker nodes onto both cloud infrastructure and volunteers' computers. However, none of the above-mentioned distributed task management systems is designed for real-time interaction. LiveQ~\cite{LiveQ} middleware platform supports real-time queuing, making volunteer computing resources suitable for an interactive game environment where a player's next-move depends on the outcome of their previous move computed at a volunteer computer. However,  the popular volunteer computing platform - World Community Grid uses BOINC~\cite{Boinc} middleware framework for the projects task management and distribution. The CitizenGrid allows flexibility to use any of task management and distribution frameworks from the list:  Copilot~\cite{Copilot}, LiveQ~\cite{LiveQ}, or BOINC~\cite{Boinc} and enable easy deployment by encapsulating these middleware platforms in a virtual machines, which not only helps developers and scientists but also volunteers as they have more flexibility to choose from their computing resources. 

\section{Conclusions and Future Work}\label{sec:conclusions}
In this paper, we have presented the CitizenGrid,  a middleware platform for online citizen science  projects. We have provided a detailed description of the platform's functionality, implementation and deployment options. The CitizenGrid  provides hosting and deployment of citizen science applications and aims to facilitate improved volunteer participation for the applications it hosts. The platform has been tested and evaluated by HCI (Human-Computer Interaction) academic evaluators. The evaluation parameters were grouped into the following categories: functionality, usability, compatibility, security, database integrity, and performance.  The current version of CitizenGrid provide an interface for running and controlling citizen science servers on Amazon EC2 and OpenStack, in future we would like to provide support for other cloud providers. The volunteer computing projects' client virtual image runs on VirtualBox on volunteer's computer, which provides safe  and secure execution environment. In future, we would like to support lightweight containerisation (Docker containers) based application deployment.

\section*{Acknowledgment}
This research was funded by the EU project Citizen Cyberlab (Grant No 317705). The authors would like to thank Jacinta M. Smith, Christine Simpson  for helping in the project.
\bibliography{pyadav_2017.bbl}

\end{document}